\documentclass[twocolumn]{aastex63}
\usepackage{fontawesome}
\usepackage{color}
\usepackage{graphicx}
\usepackage{amsmath, amssymb, bm}
\usepackage{CJK}
\usepackage[version=4]{mhchem} % chem package
\usepackage{multirow}
\submitjournal{AAS Journals}

\shorttitle{Internal Structure of GJ 1214 b}
\shortauthors{Nixon et al.}

\begin{document}
%\begin{CJK*}{UTF8}{gbsn}

\title{New insights into the internal structure of GJ 1214 b informed by JWST}
%Something like "New insights into the internal structure of GJ 1214 b informed by JWST"
%"The internal structure of GJ 1214 b: new insights informed by JWST"

\correspondingauthor{Matthew C. Nixon}
\email{mcnixon@umd.edu}

\author[0000-0001-8236-5553]{Matthew C.\ Nixon}
\affiliation{Department of Astronomy, University of Maryland, College Park, MD, USA}

\author[0000-0002-8518-9601]{Anjali A.\ A.\ Piette}
\affiliation{School of Physics and Astronomy, University of Birmingham, Edgbaston, Birmingham B15 2TT, UK}
\affiliation{Earth \& Planets Laboratory, Carnegie Institution for Science, Washington, DC, USA}

\author[0000-0002-1337-9051]{Eliza M.-R.\ Kempton}
\affiliation{Department of Astronomy, University of Maryland, College Park, MD, USA}

\author[0000-0002-8518-9601]{Peter Gao}
\affiliation{Earth \& Planets Laboratory, Carnegie Institution for Science, Washington, DC, USA}

\author[0000-0003-4733-6532]{Jacob L.\ Bean}
\affiliation{Department of Astronomy \& Astrophysics, University of Chicago, Chicago, IL, USA}

\author[0000-0001-8342-1895]{Maria E.\ Steinrueck}
\affiliation{Department of Astronomy \& Astrophysics, University of Chicago, Chicago, IL, USA}

\author[0009-0000-8049-3797]{Alexandra S.\ Mahajan}
\affiliation{Center for Astrophysics, Harvard \& Smithsonian, 60 Garden Street, Cambridge, MA 02138, USA}

\author[0000-0003-3773-5142]{Jason D.\ Eastman}
\affiliation{Center for Astrophysics, Harvard \& Smithsonian, 60 Garden Street, Cambridge, MA 02138, USA}

\author[0000-0002-0659-1783]{Michael Zhang}
\affiliation{Department of Astronomy \& Astrophysics, University of Chicago, Chicago, IL, USA}

\author[0000-0003-0638-3455]{Leslie A.\ Rogers}
\affiliation{Department of Astronomy \& Astrophysics, University of Chicago, Chicago, IL, USA}

\begin{abstract}
Recent JWST observations of the sub-Neptune GJ~1214~b suggest that it hosts a high-metallicity ($\gtrsim$100$\times$ solar), hazy atmosphere. Emission spectra of the planet show molecular absorption features, most likely due to atmospheric H$_2$O. In light of this new information, we conduct a thorough reevaluation of the planet's internal structure. We consider interior models with mixed H/He/H$_2$O envelopes of varying composition, informed by atmospheric constraints from the JWST phase curve, in order to determine possible bulk compositions and internal structures. Self-consistent atmospheric models consistent with the JWST observations are used to set boundary conditions for the interior. We find that a total envelope mass fraction of at least 8.1\% is required to explain the planet's mass and radius. Regardless of H$_2$O content, the maximum H/He mass fraction of the planet is 5.8\%. We find that a 1:1 ice-to-rock ratio along with 3.4--4.8\% H/He is also a permissible solution. In addition, we consider a pure H$_2$O (steam) envelope and find that such a scenario is possible, albeit with a high ice-to-rock ratio of at least 3.76:1, which may be unrealistic from a planet formation standpoint. We discuss possible formation pathways for the different internal structures that are consistent with observations. Since our results depend strongly on the atmospheric composition and haze properties, more precise observations of the planet's atmosphere would allow for further constraints on its internal structure. This type of analysis can be applied to any sub-Neptune with atmospheric constraints to better understand its interior.
\end{abstract}


\section{Introduction}\label{sec:intro}

One of the most important goals of modern exoplanet science is to better understand the nature of the large population of planets with radii between that of Earth and Neptune, often referred to as ``sub-Neptunes''. With no analogues in the solar system, very little prior information regarding the properties of such planets is available. Demographic studies of sub-Neptunes orbiting FGK stars indicate that this population consists of two distinct classes of planet, divided by radius \citep{Fulton2017,VanEylen2018}. The group with larger radii ($\gtrsim 1.8 R_{\oplus}$) are theorised to host substantial H/He-rich envelopes of up to a few percent by mass \citep[e.g.,][]{Owen2013}. In contrast, the population of sub-Neptunes orbiting M dwarfs appears to be separated by density rather than radius, and includes several planets of intermediate density between the two classes established for FGK stars, leading to suggestions of an additional sub-population of H$_2$O-rich ``water worlds'' orbiting such stars \citep{Luque2022}.

GJ~1214~b was one of the first sub-Neptunes to be discovered \citep{Charbonneau2009}. Since then, the planet has been studied extensively, with numerous efforts undertaken to characterise both its atmosphere and interior. Initial spectroscopic observations at optical and near-infrared wavelengths \citep{Bean2010,Bean2011,Desert2011,deMooij2012,Berta2012,Teske2013} yielded a flat, featureless transmission spectrum, indicating that the planet's atmosphere could not be both cloud-free and hydrogen-rich, but must possess either high-altitude aerosols and/or a high mean molecular weight (MMW). Extensive follow-up in the near-infrared with the Hubble Space Telescope \citep{Kreidberg2014} found that a high MMW alone could not explain the planet's transmission spectrum, meaning it must host high-altitude aerosols, but was still unable to constrain the composition.

Several internal structure modelling studies have also attempted to determine the planet's bulk composition. The bulk properties of the planet and host star are shown in Table \ref{tab:planet}. The relatively low bulk density of the planet (2.26 $\pm$ 0.18 g cm$^{-3}$) means that it likely possesses a significant gaseous component consisting of H/He and/or other chemical species. \citet{Rogers2010b} explored a range of internal structure models that could explain this bulk density, but degeneracies between model solutions made it impossible to infer a unique composition. They also found that the planet would be too hot to host liquid water. Evolutionary models from \citet{Nettelmann2011} indicated that the planet was more likely to host a mixed H/He/H$_2$O envelope than to be a pure water world with an entirely steam atmosphere, since a pure H$_2$O atmosphere would require an extremely high ice-to-rock ratio of $\sim$6 to 1 to explain the planet's mass and radius. \citet{Valencia2013} also noted that it was not possible to distinguish between a pure H/He, pure H$_2$O, or mixed envelope, but did place an overall upper limit of $\sim$7\% on the H/He mass fraction of the planet. A later study \citep{Cloutier2021} revised the mass of the planet and placed new constraints on the H/He mass fraction in the case of a pure H/He envelope ($x_{\rm env} = 5.24^{+0.30}_{-0.29}\%$).

Further information regarding the nature of GJ~1214~b's atmosphere was revealed thanks to observations of its thermal emission phase curve obtained with the mid-infrared instrument's low-resolution spectrometer (MIRI/LRS) on board JWST \citep{Kempton2023}. These observations ruled out a pure H/He envelope for the planet, since models with a high metallicity ($\gtrsim$100$\times$ solar) were required to explain the large amplitude of the phase curve. Furthermore, models with high-albedo hazes also yielded better fits to the observations, since clear atmospheres would absorb too much stellar radiation and were therefore globally too hot to explain the phase curve. The planet's dayside and nightside MIRI/LRS emission spectra both show $>3\sigma$ evidence of absorption features, with atmospheric retrievals indicating that H$_2$O is the most likely cause in both spectra. Additionally, a comparison of the 1--10~$\mu$m transmission spectrum with photochemical haze models \citep{Gao2023} showed that a range of haze prescriptions could explain the flat spectrum, favouring higher metallicities ($\gtrsim$300$\times$ solar) or a pure H$_2$O atmosphere.

\begin{figure}
\centering
\includegraphics[width=\columnwidth,trim={0 0 0 0},clip]{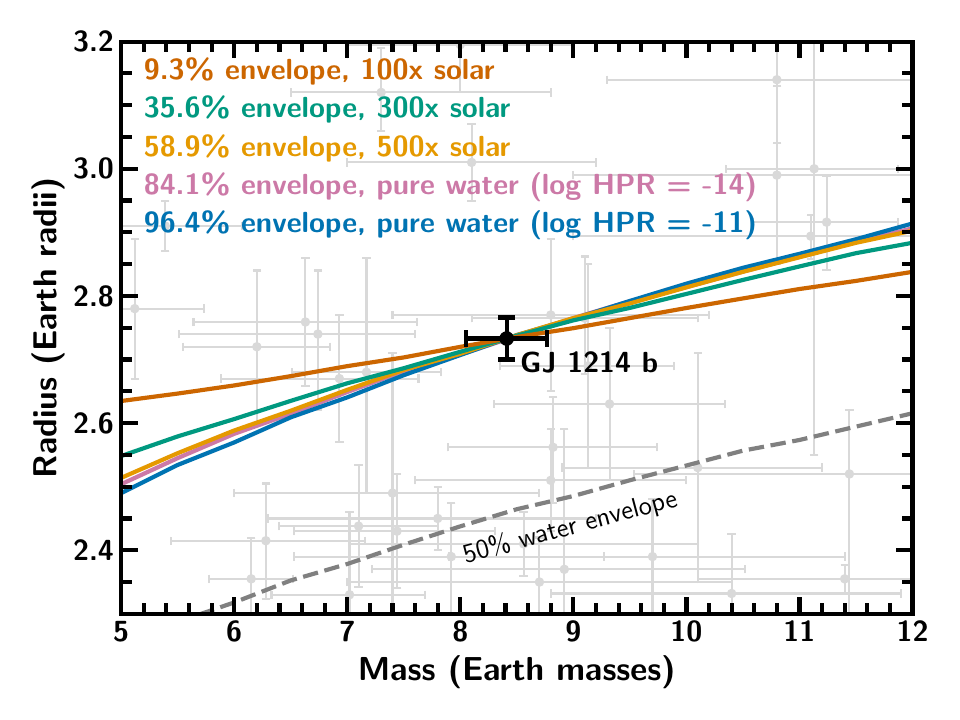}
    \caption{Mass and radius of GJ~1214~b from \citet{Mahajan2024}. Mass-radius relations for best-fit models with a selection of atmospheric compositions are shown, assuming maximally reflective hazes. The 100$\times$, 300$\times$, and 500$\times$ solar composition atmospheres assume a haze production rate (HPR) of $10^{-11}$ g cm$^{-2}$ s$^{-1}$. For the pure water (steam) atmospheres, two values of the HPR are shown: $10^{-11}$ g cm$^{-2}$ s$^{-1}$ and $10^{-14}$ g cm$^{-2}$ s$^{-1}$. Percentages shown are the total envelope mass fractions; the relevant H/He and H$_2$O mass fractions for each case can be calculated using the mass mixing ratios in Table \ref{tab:mmw}. A steam atmosphere scenario with 50\% H$_2$O and 50\% iron+rock is included for comparison, corresponding to the water world interior structure with no H/He from \citet{Luque2022}. Grey points with error bars show planets with similar masses and radii that have been measured to better than 20\% precision. Data taken from the NASA Exoplanet Archive.}
    \label{fig:mr_curves}
\end{figure}

The new insight into the atmosphere of GJ~1214~b provided by \citet{Kempton2023} and \citet{Gao2023} motivate an updated investigation of the possible internal structures of the planet. Additional impetus is provided by the work of \citet{Mahajan2024}, who used the JWST transit and secondary eclipse light curves for an improved characterization of the host star, leading to revisions of the mass and radius of the planet ($M_p=8.41^{+0.36}_{-0.35}\,M_{\oplus}$, $R_p = 2.733 \pm 0.033\,R_{\oplus}$, see Table~\ref{tab:planet} and Figure~\ref{fig:mr_curves}). In this study, we wish to determine how the incorporation of a hazy, metal-enriched atmosphere affects internal structure models. The atmospheric composition and haze properties will impact the thermal structure of the outer envelope, which can significantly impact mass-radius relations for planets with a substantial volatile component \citep{Thomas2016,Nixon2021}. We therefore combine self-consistent atmospheric models with internal structure models to place updated constraints on the bulk composition of the planet.

In Section \ref{sec:methods} we describe the internal structure model used to characterise GJ~1214~b in this study, as well as the self-consistent atmospheric model used to find the temperature profile of the outer envelope. We present constraints on the internal structure from these models in Section \ref{sec:results}, and discuss the implications of our findings, as well as caveats and future directions of study, in Section \ref{sec:discussion}.

\begin{table}
    \centering
    \setlength{\arrayrulewidth}{1.3pt}
    \begin{tabular}{ccc}
    	\hline
		Quantity (unit) & Value & Source \\
		\hline
        $R_p$ ($R_{\oplus}$) & 2.733 $\pm$ 0.033 & \citet{Mahajan2024} \\
        $M_p$ ($M_{\oplus}$) & 8.41$^{+0.36}_{-0.35}$ & \citet{Mahajan2024} \\
        $\rho_p$ (g cm$^{-3}$) & 2.26 $\pm$ 0.18 & Derived \\
        $T_{\rm eq}$ (K) & 596 $\pm$ 19 & \citet{Cloutier2021} \\
        $R_*$ ($R_{\odot}$) & 0.2162$^{+0.0025}_{-0.0024}$ & \citet{Mahajan2024} \\
        $M_*$ ($R_{\odot}$) & 0.1820$^{+0.0042}_{-0.0041}$ & \citet{Mahajan2024} \\
        $T_{\rm eff}$ (K) & 3250 $\pm$ 100 & \citet{Cloutier2021} \\
        $[$Fe/H$]_*$ (dex) & 0.24 $\pm$ 0.11 & \citet{Mahajan2024} \\
        \hline
    \end{tabular}
    \caption{Planetary and stellar properties of GJ~1214~b and its host star. Planetary equilibrium temperature $T_{\rm eq}$ is calculated assuming full heat redistribution and zero Bond albedo.}
    \label{tab:planet}
\end{table}

\newpage

\section{Methods}\label{sec:methods}

\begin{figure*}
\centering
\includegraphics[width=\textwidth,trim={0 0 0 0},clip]{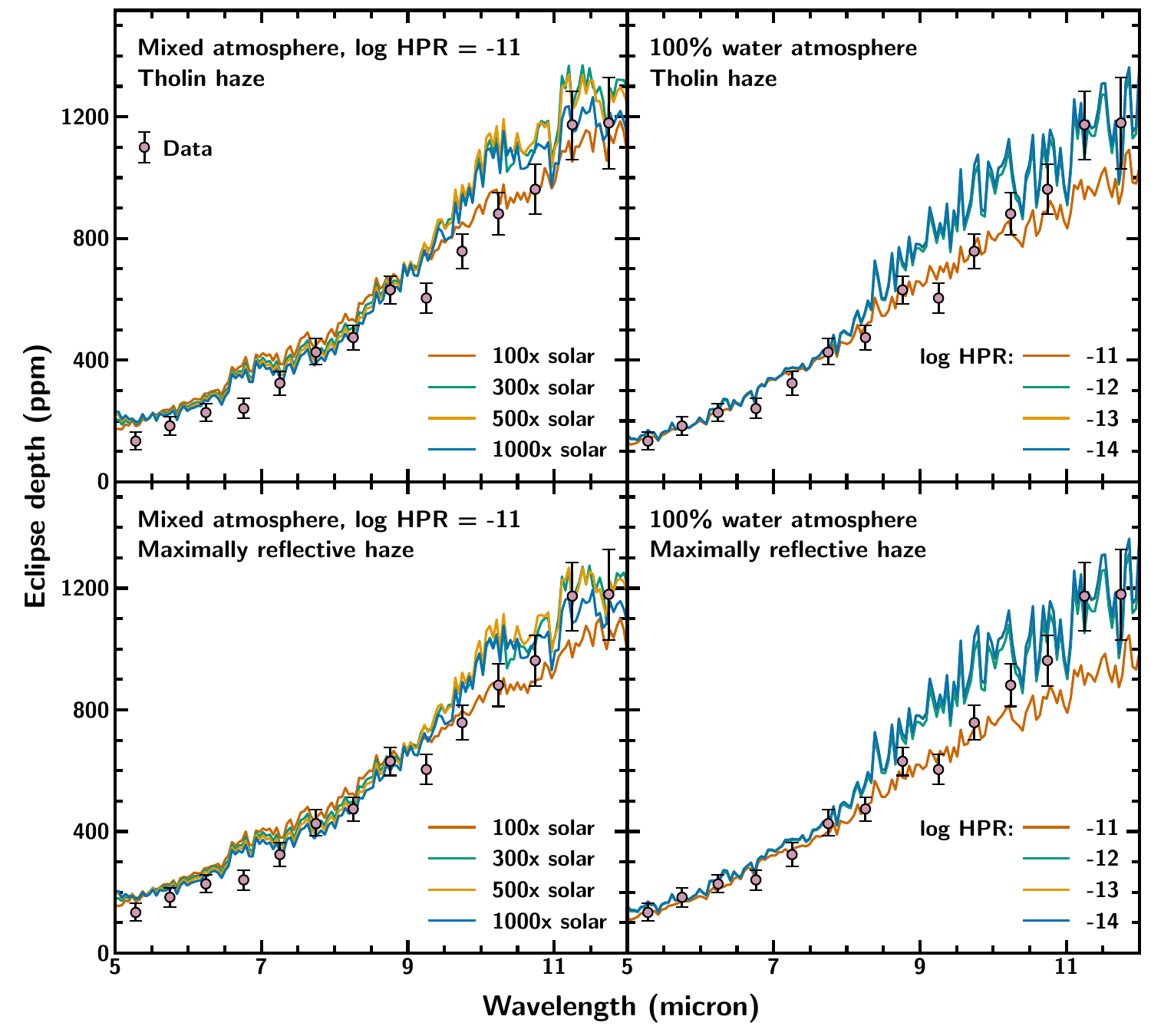}
    \caption{Model emission spectra from GENESIS with different compositions and haze production rates (HPR), compared to the MIRI dayside spectrum of GJ~1214~b. The top panels show models including Titan-like tholin hazes, while the bottom panels show models with maximally reflective, purely scattering hazes. Models with the smallest values of $\chi^{2}/n_{\rm data}$ have been chosen from a larger grid of compositions and haze production rates. Note that the water atmosphere models with HPRs of 10$^{-13}$ and 10$^{-14}$~g~cm$^{-2}$~s$^{-1}$ overlap significantly.}
    \label{fig:emission_spectra}
\end{figure*}

We characterise the internal structure of GJ~1214~b using an updated version of the model described in \citet{Nixon2021}, which has been used to characterise several other sub-Neptunes \citep{Madhu2020,Luque2021,Luque2022_G940b}. Our model solves the planetary structure equations of mass continuity and hydrostatic equilibrium assuming spherical symmetry. Model planets consist of iron, silicates (namely MgSiO$_3$), water and H/He. Throughout the text we refer to the iron and silicate component as the ``nucleus'' and the water and H/He component as the ``envelope''. We use the term ``atmosphere'' to refer to the outer region of the envelope ($P \leq 100$ bar). The equation of state (EOS) prescription for the iron core is adopted from \citet{Seager2007}, who used a Vinet EOS of the $\epsilon$ phase of Fe \citep{Vinet1989,Anderson2001}. We use an isothermal EOS for iron, since thermal effects in the core should not significantly affect the planetary radius \citep{Howe2014}. We use an updated, temperature-dependent silicate EOS following the phase diagram shown in \citet{Huang2022} and including three phases: bridgmanite, post-perovskite and liquid MgSiO$_3$. We use a Vinet EOS from \citet{Oganov2004} for bridgmanite, a different Vinet EOS from \citet{Sakai2016} for post-perovskite, and an EOS following the RTPress formulation \citep{Wolf2018} for liquid MgSiO$_3$. We also use a temperature-dependent EOS for the outer H$_2$O and H/He layers, noting that the temperature profile of the outer envelope can strongly impact the mass-radius relation \citep{Thomas2016}. For H$_2$O, we apply the EOS compiled in \citet{Nixon2021}, which was constructed from a range of sources \citep{Salpeter1967,Fei1993,Wagner2002,Feistel2006,Seager2007,French2009,Klotz2017,Journaux2020_EOS} in order to cover the full pressure-temperature ($P$--$T$) space relevant to sub-Neptune interiors. The H/He EOS is taken from \citet{Chabrier2019}, which assumes a helium mass fraction $Y=0.275$.

Since the temperature profile of a planet can have a substantial effect on its internal structure, it is important to consider realistic $P$--$T$  profiles when analysing a planet's interior. Previous studies of the internal structure of GJ~1214~b used analytic $P$--$T$ profiles calculated using the double gray approximation \citep{Rogers2010b}. Given that recent JWST observations suggest that GJ~1214~b has a hazy atmosphere, we wish to include the effect of hazes on the planet's temperature profile in the present study, an effect which was not considered in previous analyses of the planet's interior. We therefore calculate hazy dayside $P$--$T$ profiles for GJ~1214~b using an adaptation of GENESIS \citep{Gandhi2017} suited to mini-Neptune atmospheres \citep{Piette2020}. GENESIS is a self-consistent 1D atmospheric model that computes equilibrium $P$--$T$ profiles and thermal emission spectra under the assumptions of radiative–convective, hydrostatic, and local thermodynamic equilibrium.

We consider a range of $P$--$T$ profiles, calculated using different envelope compositions and haze properties informed by recent JWST observations \citep{Kempton2023,Gao2023}. Specifically, we use the following compositions as inputs for the self-consistent models: 100$\times$ solar, 300$\times$ solar, 500$\times$ solar, 1000$\times$ solar, and a steam atmosphere (100\% H$_2$O). The model includes opacity due to  H$_2$O, CH$_4$, NH$_3$, CO, CO$_2$, HCN and C$_2$H$_2$, which are the dominant opacity sources expected in hydrogen-rich atmospheres \citep{Madhu2012_CtoO,Moses2013b}. For the $N\times$ solar compositions, the abundance of each chemical species is calculated self-consistently with the temperature profile using \texttt{FastChem 2} \citep{Stock2018,Stock2022}, assuming a solar carbon-to-oxygen ratio (C/O) of 0.54 \citep{Asplund2009}. We additionally include opacity due to H$_2$-H$_2$ and H$_2$-He collision-induced absorption (CIA). The absorption cross sections of each of these opacity sources are calculated using the methods of \citet{Gandhi2017}, and data from the ExoMol, HITRAN and HITEMP databases (H$_2$O, CO and CO$_2$: \citealt{Rothman2010}, CH$_4$: \citealt{Yurchenko2013,Yurchenko2014a}, C$_2$H$_2$: \citealt{Rothman2013,Gordon2017}, NH$_3$: \citealt{Yurchenko2011}, HCN: \citealt{Harris2006,Barber2014}, CIA: \citealt{Richard2012}).

The calculated $P$--$T$ profiles from GENESIS extend to a pressure of 100 bar, beyond which the profile is extended to higher pressures following an adiabat (isentrope). We assume an internal heat flux equivalent to an intrinsic temperature $T_{\rm int} = 30$ K, as adopted by \citet{Gao2023}, and which is consistent with evolutionary models of GJ~1214~b \citep{Valencia2013,Lopez2014}. We additionally assume efficient redistribution of energy between the dayside and nightside. This provides a globally-averaged temperature profile, which can be used as a boundary condition for the 1D interior structure model. However, we note that the phase curve observations of \citet{Kempton2023} suggest somewhat inefficient day-night heat redistribution, which may result in minor asymmetries between the dayside and nightside boundary conditions that are beyond the scope of this work.

In order to explain the emission spectra of GJ~1214~b presented in \citet{Kempton2023}, aerosols present in the atmosphere require a single scattering albedo at least comparable to that of Titan tholins, and potentially as high as 1 (i.e., maximally reflective). Following \citet{Gao2023}, we consider these two haze scenarios as end-member cases for our atmospheric models, where the maximally reflective hazes have refractive indices $n=1.8, k=10^{-9}$. These are identical to the ``purely scattering'' hazes in \citet{Gao2023}. Haze particle size distributions are also identical to \citet{Gao2023}. We allow the haze production rate to vary between 10$^{-14}$ and $10^{-9}$ g cm$^{-2}$ s$^{-1}$, with the lower limit set by haze production on Titan \citep{Checlair2016} and the upper limit set by estimates from photochemical models of GJ~1214~b \citep{Kawashima2019,Lavvas2019}.

For all atmospheric models considered, the $P$--$T$ profile remains hot enough that H$_2$O never enters any liquid or ice phases, instead remaining in vapour, supercritical, and superionic phases throughout the envelope. The miscibility of hydrogen and water does not appear to have been tested directly throughout the full $P$--$T$ range of these phases. \citet{Soubiran2015} found that H and H$_2$O are likely to be miscible at temperatures between 2000 and 6000~K and pressures at 10s of GPa using ab initio simulations, and \citet{Gupta2024} found H and H$_2$O to be miscible at temperatures down to 650~K at $\sim$0.01~GPa. However, some experimental studies have suggested that demixing may occur at certain pressures and temperatures \citep[e.g.,][]{Bali2013,Vlasov2023}. Studies of other chemical species indicate that gases and supercritical fluids are completely miscible \citep{Budisa2014,Pruteanu2017}, and the assumption of the miscibility of hydrogen and supercritical water has been made in previous studies of sub-Neptune interiors \citep{Pierrehumbert2023,Benneke2024}. We therefore assume that H/He and H$_2$O remain fully mixed throughout the envelope, but note that further experimental and theoretical work to test this assumption would be extremely valuable for sub-Neptune internal structure modelling.

%For a given composition, we fix the H$_2$O mass fraction such that the mean molecular weight of the envelope matches the mean molecular weight of the atmosphere calculated by the self-consistent model at 1 bar.

%Under the assumption of an Earth-like nucleus (1/3 iron, 2/3 silicates by mass) and a fully mixed H/He-H$_2$O envelope in proportions set by the chosen atmospheric composition, our model reduces to two components for a given scenario. Furthermore, since the mass fractions of each component must sum to unity, for a given atmospheric composition we only need to explore a single parameter ($x_{\rm nuc}$ or $x_{\rm env}$) in order to find the range of bulk compositions consistent with the mass and radius of GJ~1214~b. %This might go in the results section

%*UPDATE THIS* We explore the parameter space of possible compositions in ($x_{\rm nuc}$, $x_{\rm H_2O}$, $x_{\rm H/He}$) space. For each composition, we considered a range of masses that agree with the observed mass of the planet to within 1$\sigma$. For a given mass $\hat{M}$, the model radius $\hat{R}$ is computed, and the $\chi^2$ statistic is calculated:
%\begin{equation}
%    \chi^2 = \frac{(M_p-\hat{M})^2}{\sigma_M^2} + \frac{(R_p-\hat{R})^2}{\sigma_R^2},
%\end{equation}
%where ($\sigma_M,\sigma_R$) are the observed uncertainties on the mass and radius of each planet.

\begin{table}
    \centering
    \setlength{\arrayrulewidth}{1.3pt}
    \begin{tabular}{ccc}
    	\hline
		Composition & MMW (g mol$^{-1}$) & H$_2$O mass fraction \\
		\hline
        100$\times$ solar & 5.2 & 0.633 \\
        300$\times$ solar & 9.8 & 0.875 \\
        500$\times$ solar & 13.4 & 0.949 \\
        Pure H$_2$O & 18.0 & 1.000 \\
        \hline
    \end{tabular}
    \caption{Atmospheric compositions, mean molecular weights (MMW) and chosen H$_2$O mass fractions for each model under consideration.}
    \label{tab:mmw}
\end{table}

\section{Results}\label{sec:results}

Our goal is to explore the range of internal structures that are consistent with both the mass and radius measurements as well as the known atmospheric properties of GJ~1214~b. We use self-consistent atmospheric models to determine which atmospheric compositions are consistent with the planet’s emission spectrum. The compositions and thermal structures of best-fitting models are then used to set the composition and thermal structure of the envelope in the interior model, under the assumption that the atmosphere/envelope is well-mixed. This allows us to constrain the bulk composition of the planet.

We begin by comparing our grid of self-consistent atmospheric models from GENESIS to the observed dayside emission spectrum of the planet, in order to assess which compositions and haze properties are consistent with observations. Figure \ref{fig:emission_spectra} shows emission spectra from the grid which most closely fit the observations. For both tholin hazes and maximally reflective hazes we find that, across all compositions, models with a haze production rate of 10$^{-11}$g cm$^{-2}$ s$^{-1}$ yield the best fit to the spectrum. Additionally, for a pure H$_2$O atmosphere, lower haze production rates also provide a reasonable fit. We therefore consider $P$--$T$ profiles from this subset of the model grid when generating internal structure models, exploring how varying the atmospheric properties alters the possible bulk compositions consistent with the planet's mass and radius. More detailed modelling of the emission spectrum of GJ~1214~b will be presented in a future study. We note that although models with metallicities as low as 100$\times$ solar are included in our chosen subset of models based on the fit to the emission spectrum, a metallicity of $>$300$\times$ solar is preferred by the planet's transmission spectrum \citep{Gao2023}. In this study we consider metallicities ranging from 100$\times$ solar and higher, but we note that models with metallicities $>$300$\times$ solar are the most plausible in light of other observations.

\begin{figure*}
\centering
\includegraphics[width=\columnwidth,trim={0 0 0 0},clip]{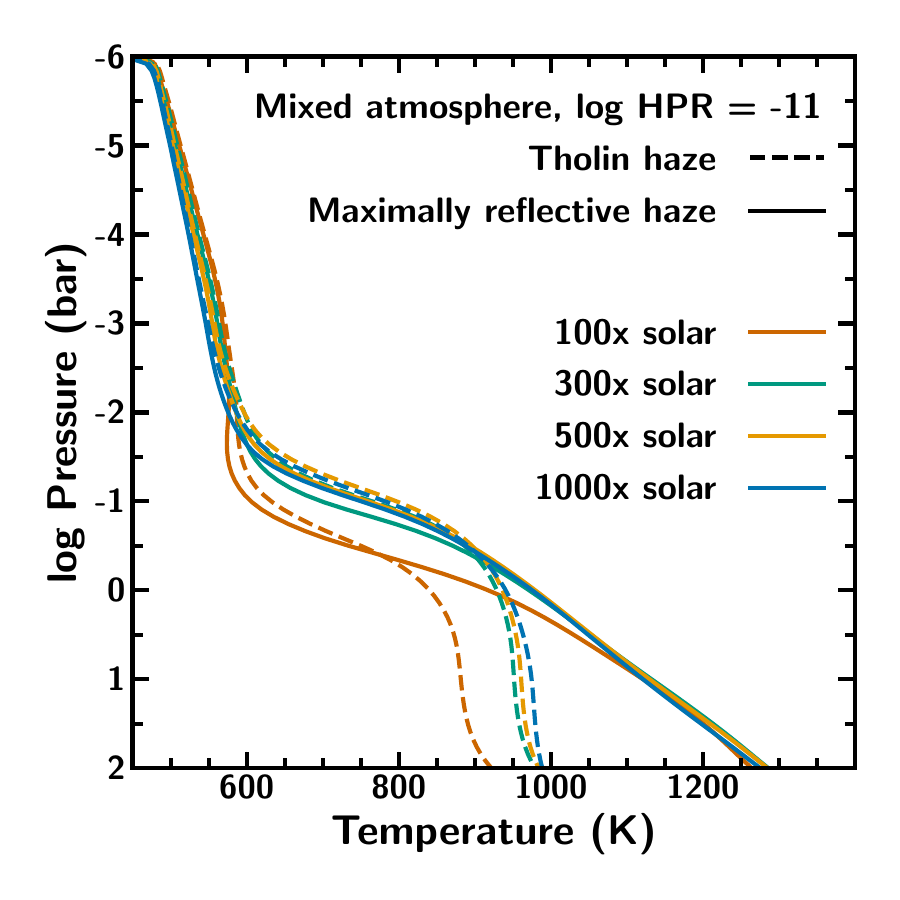}
\includegraphics[width=\columnwidth,trim={0 0 0 0},clip]{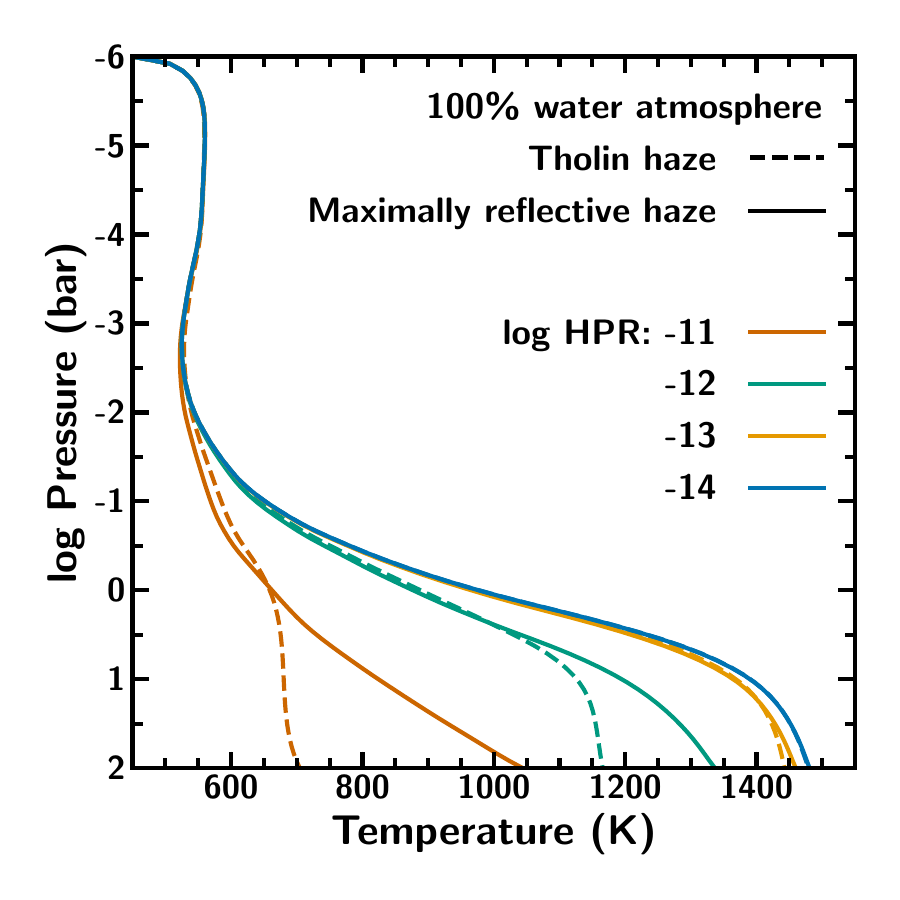}
    \caption{$P$--$T$ profiles of best-fitting GENESIS models to the emission spectrum of GJ~1214~b. The left hand panel shows models with a fixed haze production rate (10$^{-11}$g cm$^{-2}$ s$^{-1}$) and varying compositions, while the right hand panel shows profiles for a pure steam atmosphere with different haze production rates. These profiles are used to set the thermal structure of the outer envelope for the interior model grid.}
    \label{fig:hazy_pt_profiles}
\end{figure*}

We subsequently turn to internal structure models. We consider two possible scenarios for the gaseous envelope: a mixture of H/He/H$_2$O and a pure H$_2$O (steam) atmosphere. For mixed atmospheres, we choose H$_2$O mass fractions in the envelope, $x_{\rm H_2O, env}$ in order to match a given atmospheric MMW $\mu_{\rm atm}$ using the following formula:

\begin{align}
    x_{\rm H_2O, env} = \frac{\mu_{\rm H_2O} \left( \mu_{\rm atm} - \mu_{\rm H/He} \right)}{\mu_{\rm atm} \left( \mu_{\rm H_2O} - \mu_{\rm H/He} \right)},
\end{align}
where $\mu_{\rm H_2O}=18.02\,$amu and $\mu_{\rm H/He}=2.34\,$amu represent the MMW of H$_2$O and H/He (where H and He are mixed with a helium mass fraction $Y=0.275$). Table \ref{tab:mmw} shows the MMW and H$_2$O mass mixing ratios for metallicities included in the atmospheric model grid. We compute the thermal structure by interpolating between the $P$--$T$ profiles shown in Figure \ref{fig:hazy_pt_profiles} with the closest metallicities. For the H/He/H$_2$O envelopes we consider MMW values from 5--17 amu, since lower values would correspond to atmospheric metallicites $\lesssim$100$\times$ solar, which are unlikely given the observations. 

Under the assumption of an Earth-like nucleus (1/3 iron, 2/3 silicates by mass), and that an H/He/H$_2$O envelope would remain fully mixed, only a single free parameter ($x_{\rm env}$) remains for a given atmospheric composition. We note that while other chemical species such as CO$_2$ may be present in the atmosphere, we choose a composition of H/He/H$_2$O partly since there is observational evidence for H$_2$O absorption in the planet's emission spectra, and partly because there is a lack of high-pressure EOS data for other chemical species. However, constraints on the H/He mass fraction derived here are unlikely to change regardless of whether the main non-H/He constituent of the atmosphere is H$_2$O or a combination of other chemical species (see Section \ref{sec:discussion} for more detail).

Figure \ref{fig:mr_curves} shows the mass and radius of GJ~1214~b alongside a selection of best-fitting models incorporating maximally reflective hazes. We can see from this figure that it is possible to explain the mass and radius of the planet with any of the envelope compositions we consider, due to inherent degeneracies present in internal structure models. However, the best-fitting envelope mass fractions vary significantly between different compositions. We therefore explore in detail how the permitted envelope mass fraction depends on its composition.

\begin{figure*}
\centering
\includegraphics[width=0.49\textwidth,trim={0 0 0 0},clip]{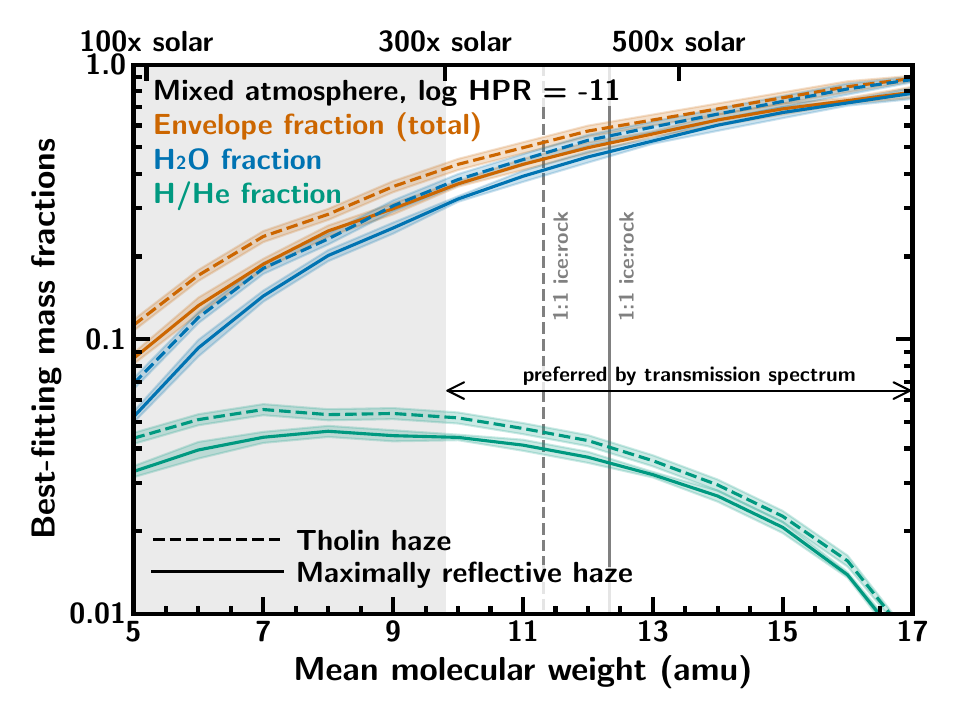}
\includegraphics[width=0.49\textwidth,trim={0 0 0 0},clip]{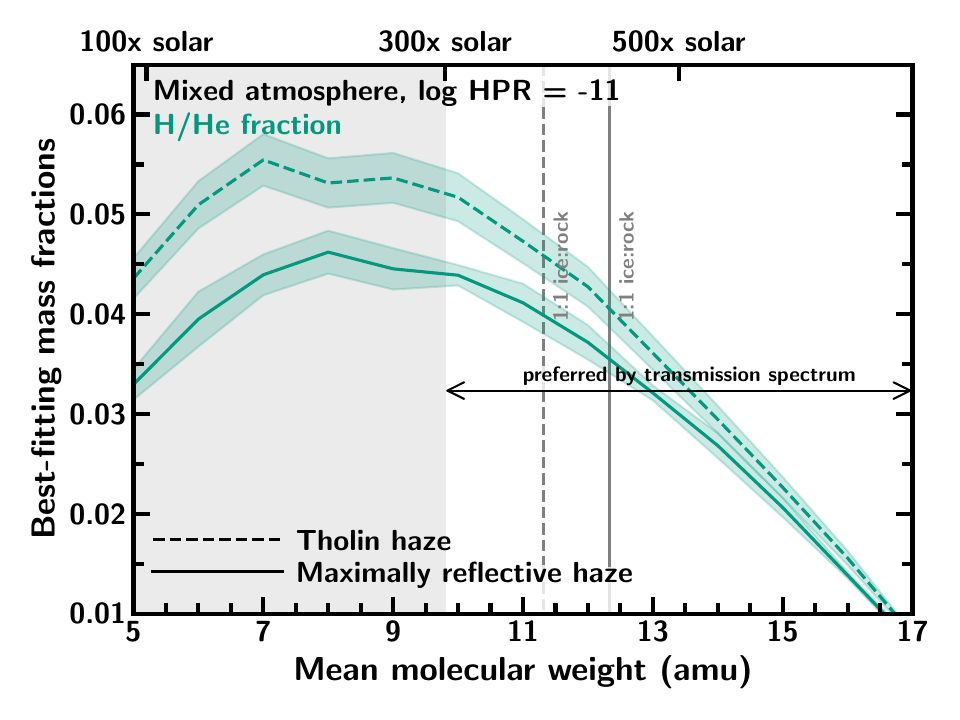}
    \caption{\textit{Left:} best-fitting envelope mass fractions for GJ~1214~b as a function of envelope mean molecular weight. Solid lines  (solid and dashed lines). Shaded regions around each line correspond to values that are consistent with the observed mass and radius to within 1$\sigma$. The total envelope fraction is shown in orange (uppermost lines), while the corresponding H$_2$O and H/He fractions are shown in blue (middle lines) and green (bottom lines) respectively. The planet's phase curve suggests that the atmospheric metallicity is at least 100$\times$ solar, equivalent to a total envelope mass fraction of at least 8.1\%, while the transmission spectrum prefers metallicities $\geq$300$\times$ solar, yielding a minimum envelope fraction of 34.4\%. The region of the figure showing models with metalliticies $<$300$\times$ solar is shaded in grey. {Right:} same as left panel, but only showing H/He fraction on a linear scale for clarity. The maximum H/He mass fraction across all models considered is 5.8\%.}
    \label{fig:best_env}
\end{figure*}

\begin{figure*}
\centering
\includegraphics[width=\textwidth,trim={0 0 0 0},clip]{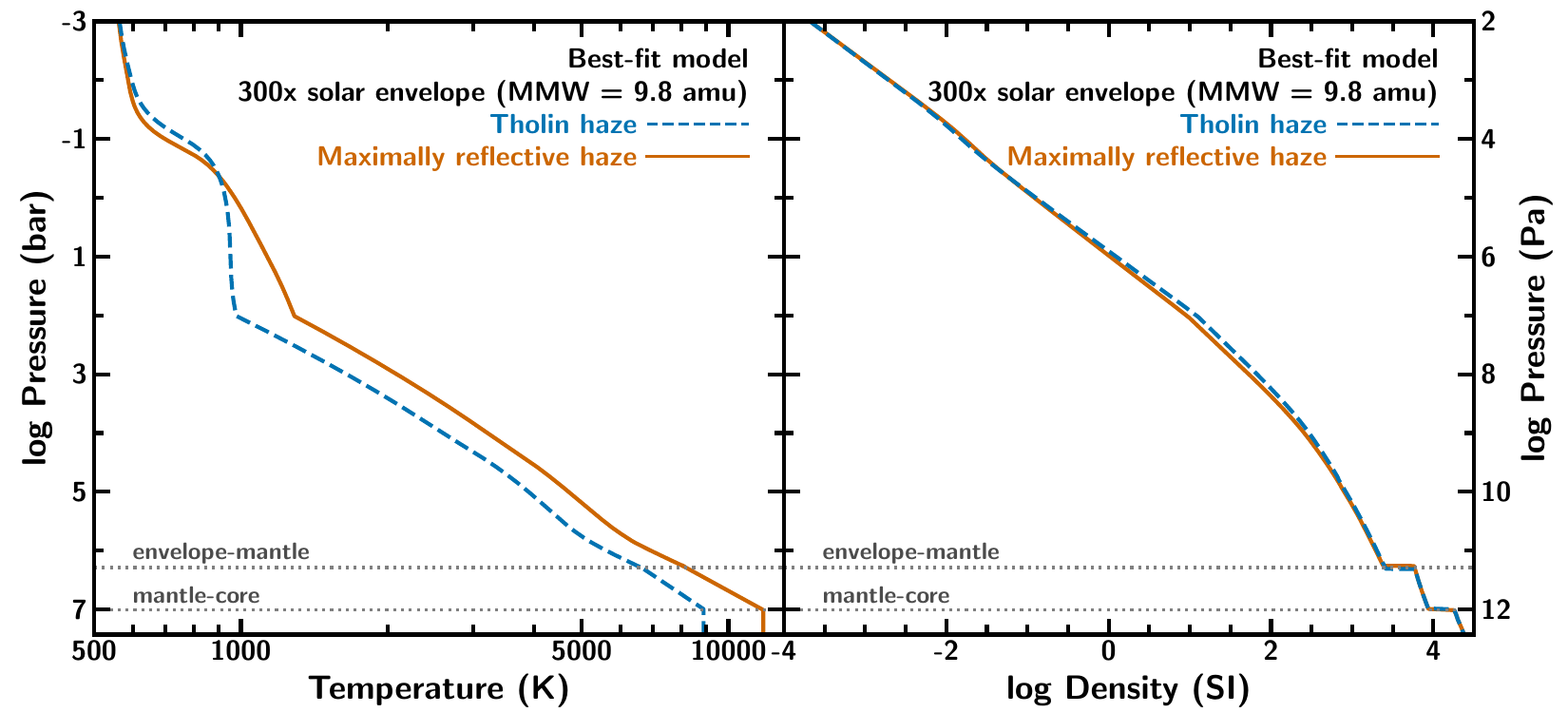}
    \caption{Interior $P$--$T$ (left) and $P$--$\rho$ (right) profiles for best-fit models with a 300$\times$ solar composition envelope. Since the maximally reflective hazes lead to a hotter envelope, the envelope density is lower, leading to smaller envelope mass fractions that are required to match the mass and radius of GJ~1214~b. Note that we do not calculate a temperature gradient for the core, since our EOS for iron is not temperature-dependent.}
    \label{fig:ptrho}
\end{figure*}

Resulting envelope mass fractions that fit the mass and radius of GJ~1214~b to within 1$\sigma$ for models with a mixed H/He/H$_2$O envelope and a fixed haze production rate of 10$^{-11}$g cm$^{-2}$ s$^{-1}$ are shown in Figure \ref{fig:best_env}. We find that, as the atmospheric metallicity increases, so does the required envelope mass fraction, since a more massive envelope is required to allow the denser material to account for the same volume. Likewise, across all atmospheric compositions, models with tholin hazes require a higher envelope mass fraction than those with maximally reflective hazes in order to explain the planet's mass and radius. Best-fitting models with tholin hazes have envelopes which are up to 1.55$\times$ more massive than those with maximally reflective hazes. This is due to the tholins having higher UV extinction than the maximally reflective hazes, which are purely scattering. This yields deeper isotherms, which in turn leads to cooler, and therefore denser, envelopes overall. We illustrate this effect for the 300$\times$ solar envelope case in Figure \ref{fig:ptrho}, which shows the $P$--$T$ and $P$--$\rho$ profiles for the best-fit models with both haze prescriptions.

Across all compositions considered, the envelope must be at least 8.1\% by mass. This corresponds to a H$_2$O mass fraction of 5.0\%, a H/He mass fraction of 3.1\%, and an iron+rock mass fraction of 91.9\%. This represents the maximum possible mass fraction of the iron+rock nucleus. For the higher metallicity scenarios ($\geq$300$\times$ solar) that are the most consistent with the observed phase curve and transmission spectrum, the minimum envelope mass fraction rises to 34.4\%. For the highest MMW mixed atmosphere that we consider (17 amu), the total envelope mass fraction can be up to 91.1\%. However, this solution consists almost entirely of H$_2$O (90.3\% by mass) and is likely unrealistic from a planet formation scenario, as discussed in Section \ref{sec:discussion}. One common way of setting an upper limit for the H$_2$O mass fraction is to allow the H$_2$O layer and the iron+rock nucleus to have equal mass fractions \citep[e.g.,][]{Luque2022}, which is derived from the solar system ice-to-rock ratio \citep{Lodders2003}. We find that such a solution is permissible by the data: for the tholin haze case, this corresponds to $x_{\rm H_2O} = x_{\rm nuc} = 47.6\%, x_{\rm H/He} = 4.8\%$, and in the maximxally reflective haze case, this corresponds to $x_{\rm H_2O} = x_{\rm nuc} = 48.3\%, x_{\rm H/He} = 3.4\%$.

For models with a MMW $\lesssim$8 amu, the H/He mass fraction increases as the total envelope mass fraction increases. However, at higher MMW, H$_2$O dominates over H/He in the atmosphere such that the total H/He fraction starts to decrease. This leads to an overall maximum H/He mass fraction for the planet of 5.8\% for tholin hazes (or 4.8\% for maximally reflective hazes). While the H/He fraction could technically drop as low as zero for e.g., a pure steam atmosphere, such a scenario would require a very high H$_2$O mass fraction, as discussed below. Considering models preferred by the transmission spectrum ($\geq$300$\times$ solar metallicity) with an ice-to-rock ratio $\leq$1, the H/He mass fraction must range between 3.4--5.2\%.

\begin{figure}
\centering
\includegraphics[width=\columnwidth,trim={0 0 0 0},clip]{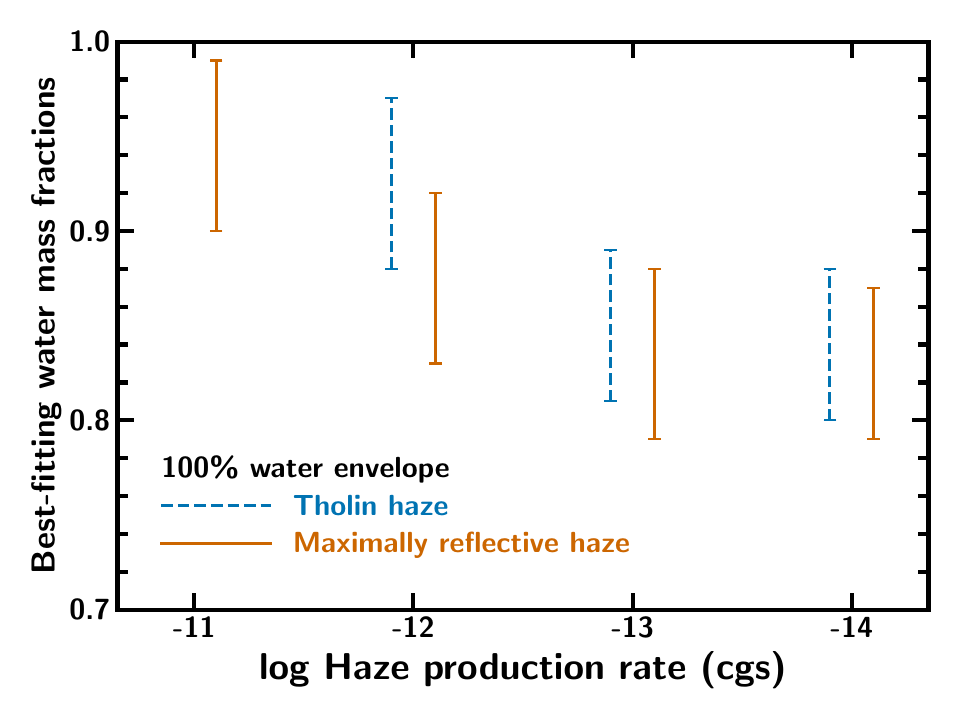}
    \caption{H$_2$O mass fractions consistent with the observed mass and radius of GJ~1214~b to within 1$\sigma$, in the case of a pure water envelope, for different haze production rates. Across all cases, the H$_2$O mass fraction  must be at least 79\%. For a maximal haze production rate (10$^{-11}$g~cm$^{-2}$~s$^{-1}$) of tholin hazes, even a 100\% H$_2$O mass fraction is not sufficient to match the planet's mass and radius.}
    \label{fig:best_env_h2o}
\end{figure}

Figure \ref{fig:best_env_h2o} shows best-fitting H$_2$O mass fractions in the case of a pure steam atmosphere, with no H/He. Since lower haze production rates lead to a hotter atmosphere, the value of $x_{\rm H_2O}$ required to explain the mass and radius decreases. Furthermore, at lower haze production rates, the hazes have less of an impact on the thermal structure, meaning that the difference between the tholin haze models and the maximally reflective haze models decreases. In the tholin haze case, the highest haze production rate (10$^{-11}$g cm$^{-2}$ s$^{-1}$) cannot be fit with a pure H$_2$O envelope; even a 100\% H$_2$O mass fraction yields a planet that is too small. In contrast, the maximally reflective haze model with the same haze production rate can have a water mass fraction of 90--99\%. For the lowest haze production rates considered, the models require a water mass fraction of at least 79\%. The feasibility of such a substantial H$_2$O component from a planet formation perspective is discussed in Section \ref{sec:discussion}.

\section{Discussion and Conclusions}\label{sec:discussion}

\subsection{Implications for planet formation}

The indications of a high-metallicity atmosphere and H$_2$O absorption features found in the JWST observations of GJ~1214~b suggest that the planet may have a substantial water component. It has long been theorised that such planets could come into being through formation outside the water ice line of  protoplanetary disks and subsequent inward migration \citep[e.g.,][]{Leger2004}. In the case of GJ~1214~b, this inward migration and corresponding increase in irradiation would cause the planet's H$_2$O to evaporate into a steam/supercritical atmosphere. The resulting envelope composition would depend on whether the planet had also accreted a substantial H/He envelope at formation. If H/He accretion had taken place, the planet would end up with a mixed H/He/H$_2$O envelope, and if not, it would host a pure steam atmosphere. These two formation and evolution scenarios are therefore analogous to the two modelling cases covered in this work, and we discuss their feasibility here.

\subsubsection{Mixed hydrogen/helium/water envelope}

In this scenario, the ice/rock nucleus would accrete some amount of H/He, which would subsequently mix with the H$_2$O layer to form an extended H/He/H$_2$O envelope, with H$_2$O in vapour and/or supercritical phases. A commonly assumed composition for water worlds is a 1:1 ratio of iron and rock to H$_2$O \citep[e.g.,][]{Zeng2019,Luque2022,Chakrabarty2023}. This ratio is ultimately derived from the estimated solar system ice-to-rock ratio of 1.17:1 reported in \citet{Lodders2003}. As presented in Section \ref{sec:results}, a 1:1 ice-to-rock ratio would correspond to a H/He mass fraction of 3.4--4.8\%. Following theoretical predictions from \citet{Ginzburg2016}, we would expect GJ~1214~b to host an initial H/He mass fraction of 2.2\% after gaseous accretion and boil off. This would suggest that the above scenario is unlikely. However, we note that \citet{Ginzburg2016} only consider accretion on to a rocky, and not icy, nucleus, and that the details of the boil-off process are highly uncertain \citep{Rogers2024}. Furthermore, as can be inferred from Figure \ref{fig:best_env}, lower ice-to-rock ratios would lead to required H/He mass fractions closer to 2\%, meaning this scenario could still be possible. Relaxing the assumption of an Earth-like nucleus composition could also change the required H/He mass fraction for this scenario: for example, a purely silicate nucleus with no iron would lead to lower required H/He fractions.

An alternative pathway to forming a high-metallicity atmosphere without significant accretion of ice at formation could be the production of volatiles due to reactions between the iron+rock nucleus and the H/He-rich envelope \citep[e.g.,][]{Schlichting2022}. However, the impact of reactions at the nucleus-envelope interface on the composition of the outer envelope and atmosphere remains poorly understood.

\subsubsection{Pure water envelope}

In this case, the ice+rock nucleus would be devoid of any H/He, either due to a lack of accretion or the complete stripping of H/He during evolution. Previous work stated that this scenario would require an ice-to-rock ratio of $\gtrsim$6:1 in order to be consistent with the bulk properties of the GJ~1214~b \citep{Nettelmann2011}. Our work indicates that the planet could host a pure H$_2$O envelope with a somewhat smaller ratio of least 3.76:1. While still much higher than solar system values, similarly large H$_2$O mass fractions for sub-Neptunes have been proposed; for example, \citet{Marcus2010} consider an upper limit of 75\% for the water mass fraction at formation (an ice-to-rock ratio of 3:1). Furthermore, studies of solar system ice giants have invoked ice-to-rock ratios as high as 15 for Neptune \citep{Nettelmann2013}, though the validity of this modelling has been called into question \citep{Helled2020}. Ultimately, while it is assumed that some amount of rocky material is required to initiate further ice and gas accretion, there is no consensus on a realistic upper limit for the H$_2$O mass fraction for a sub-Neptune. While several studies have explored the available H$_2$O budget for planet formation \citep[e.g.,][]{Bitsch2021,Burn2024}, further theoretical work into the formation and evolution of icy planets is required to determine whether such a composition is truly plausible. Our work suggests that a H/He-free water world scenario is unlikely for GJ~1214~b, though if the planet is ultimately determined to have such a composition, it would have interesting implications for the reservoir of icy material available in the original disk.

\subsection{Comparison with previous studies}

Early efforts to quantify the maximum H/He mass fraction possible for GJ~1214~b both obtained an upper limit of $\sim$7\% within 1$\sigma$ \citep{Rogers2010b,Valencia2013}. The higher planet mass presented in \citet{Cloutier2021} led to a higher bulk density for the planet, which correspondingly lowered the 1$\sigma$ upper limit for $x_{\rm{H/He}}$ to 5.54\%. However, the models used to derive this upper limit assumed an isothermal atmosphere at the planet's equilibrium temperature for a solar-composition H/He envelope. The present work advances on this modeling in two important ways. First of all, we do not use an isothermal temperature profile, instead using profiles taken from self-consistent models which impose thermo-chemical and radiative-convective equilibrium as well as accounting for the effects of clouds and hazes, which have been shown to be present in GJ~1214~b's the atmosphere from JWST observations \citep{Kempton2023}. This leads to hotter $P$--$T$ profiles overall than would be obtained by assuming an isotherm at $T_{\rm eq}$. The second key difference is that we do not assume a solar composition atmosphere, instead varying the composition within a range of values consistent with the JWST data. While the hotter atmosphere leads to a more inflated envelope, thus reducing the allowed H/He mass fraction, the compositional difference has the opposite effect, meaning that larger H/He fractions are permitted. In the lowest metallicity cases presented here (100$\times$ solar), we find that lower H/He mass fractions than past work are permissible ($\sim$3\%). However, at higher metallicities, we find that H/He mass fractions slightly larger than those reported in \citet{Cloutier2021} are permissible, up to $\sim$6\%. These results strongly depend on the specific atmospheric composition that is considered.

\subsection{Caveats and model improvements}

In order to better understand the internal structures of planets much larger than Earth, we require more information on the behaviour of materials at the high pressures found in giant planet interiors. For example, experimental data on the equation of state of H$_2$O at high pressures remains limited \citep{Journaux2020}, meaning the densities used in internal structure models are based on theoretical approximations. Validating such approximations will be extremely valuable in refining sub-Neptune interior models. Additionally, while studies have shown that H/He and H$_2$O are expected to be miscible and approximately follow an ideal mixing law at certain pressures and temperatures \citep{Soubiran2015}, this must be extended to a wider range of thermodynamic conditions to determine whether the assumption of a mixed H/He-H$_2$O envelope throughout GJ~1214~b is valid, and whether the region of the atmosphere probed through observations ($P \lesssim 1$ bar) is representative of the entire envelope. If there was a compositional gradient such that the water enrichment was greater moving deeper into the planet, then we would expect the required envelope mass fractions to increase due to its higher density. In this sense, the fully mixed case presented here can be thought of as providing a lower limit to the required envelope fraction. It is also possible that additional H$_2$O may be partitioned in the iron/rock nucleus of the planet \citep{Dorn2021,Kovacevic2022,Luo2024}, an effect which must be better understood in order to refine measurements of planetary bulk water mass fractions.

In our internal structure models, we have made the assumption that the outer envelope of GJ~1214~b consists of H, He and H$_2$O, based on observational evidence as well as modelling limitations. While other chemical species may be present in the envelope, we do not possess high-pressure EOS data for these species and therefore they cannot be included in our models at present. Theoretical and experimental work to determine the high-pressure EOS of chemical species such as CO$_2$ and CH$_4$ would be invaluable to enable more detailed modelling efforts of the internal structures of sub-Neptunes.

While we have assumed an intrinsic temperature $T_{\rm int}=30\,$K for the atmospheric models used in this study, following \citet{Gao2023}, we note that there is some uncertainty in this value due to the uncertain age of the planet. We recomputed $P$--$T$ profiles with a higher value of $T_{\rm int}=40\,$K and did not find a significant change in the planetary radius. For example, for a planet with $M_p = 8.41\,M_{\oplus}$ and a 300$\times$ solar envelope with $x_{\rm env}=0.356$ and maximally reflective hazes, the radius assuming $T_{\rm int}=30\,$K is $2.727\,R_{\oplus}$, increasing to $2.729\,R_{\oplus}$ at $T_{\rm int}=40\,$K.

Internal structure modelling efforts would also benefit from a better understanding of haze microphysics. For example, it is unclear how hazes would be able to form in a pure steam atmosphere \citep{Gao2023}. Additionally, the haze models considered in this study use spherical haze particles, but porous fractal aggregates could also be present, which have an impact on resultant spectra and $P$--$T$ profiles \citep{Adams2019}, potentially allowing for lower haze production rates. However, the impact of such haze properties has not been explored in detail and is beyond the scope of this study.

\subsection{Benefit of additional observations}

The findings of this work show that JWST observations of sub-Neptune atmospheres can be used to learn about internal structures, with the composition, haze properties and thermal structure of the atmosphere inferred from GJ~1214~b's phase curve all affecting constraints on the possible bulk composition of the planet. This motivates further atmospheric observations of the planet, as well as of other sub-Neptunes. A more precise measurement of the atmospheric composition of GJ~1214~b would further limit the possible cases beyond those considered in this work, and lead to better constraints on the internal structure. Furthermore, observations of different sub-Neptunes could allow for population studies that may point towards particular formation pathways or common bulk compositions.

We have assumed that H$_2$O is the main component of the atmosphere other than H/He, since it was identified as the most likely cause of the features detected in GJ~1214~b's emission spectrum \citep{Kempton2023}. However, this detection is far from definitive, with 2.5$\sigma$ confidence on the dayside and 2.6$\sigma$ confidence on the nightside, and the constraints on the H$_2$O abundance are very broad. Furthermore, the C/O for the atmospheric models used here is assumed to be solar (C/O$=0.54$), since we have so far been unable to measure the abundance of any carbon-bearing species on the planet. Acquiring precise constraints on the atmospheric metallicity and C/O would restrict the set of internal structure models that could be applied to this planet, helping to break degeneracies and better understand the planet's bulk composition, as well as allowing the use of better-informed atmospheric temperature profiles that more accurately represent the range of relevant opacity sources. We note, however, that because our model grid encompasses a wide range of values of the envelope MMW, and the temperature of the planet means that phase transitions are not a concern, the constraints placed on the H/He mass fraction in this study are unlikely to change significantly even in the presence of additional atmospheric species other than H$_2$O.

\newpage

\subsection{Conclusions}

Recent observations of the atmosphere of GJ~1214~b indicated that the planet possesses a high-metallicity atmosphere containing reflective hazes. We have used this information alongside internal structure models and updated measurements of the planet's mass and radius \citep[$M_p=8.41^{+0.36}_{-0.35}\,M_{\oplus}$, $R_p = 2.733 \pm 0.033\,R_{\oplus}$,][]{Mahajan2024} to place constraints on the possible bulk composition of the planet. We considered models with an envelope consisting of mixed H/He/H$_2$O as well as pure H$_2$O. Across the range of models considered, we find the following:

\begin{itemize}
    \item GJ~1214~b hosts a volatile envelope of at least 8.1\% by mass. This lower limit is derived from the lowest metallicity atmosphere that remains consistent with the phase curve observations (100$\times$ solar). This limit increases to 34.4\% if only models with metallicities $\geq$300$\times$ solar are considered, as preferred by the transmission spectrum. 
    \item While a 100\% volatile envelope consisting of pure H$_2$O is consistent with the mass and radius, it is very unlikely that a planet would form with this composition.
    \item Regardless of H$_2$O content, the maximum H/He mass fraction for the planet is 5.8\%. If we require an ice-to-rock ratio of $\lesssim$1:1, then the minimum H/He mass fraction is 3.4\%.
    \item Assuming that the volatile envelope consists primarily of H, He and H$_2$O, the planet must have a total H$_2$O mass fraction of at least 5.0\%.
    \item Viable solutions exist in which the planet has H$_2$O and iron/rock components of equal mass, i.e., the ``water world'' scenario described by \citep{Luque2022}. In this case, the H/He mass fraction must be 3.4--4.8\%.
    \item If the planet has a pure steam atmosphere, with no H/He, then it must have a very high ice-to-rock ratio of at least 3.76:1. It is unclear whether this is realistic from a planet formation perspective.
    \item The atmospheric composition of a given model, including assumptions about aerosols, strongly influences the inferred bulk H/He fraction.
\end{itemize}

JWST holds the promise of enabling a much more comprehensive understanding of the atmospheres of sub-Neptunes than has previously been possible, unlocking new information about their composition, aerosol properties and thermal structure. Alongside GJ~1214~b, a number of sub-Neptunes in the 2--3$R_{\oplus}$ regime have already been observed using JWST, leading to constraints on their atmospheric composition, including K2-18~b \citep{Madhusudhan2023} and TOI-270~d \citep{Benneke2024}, with many more targets set to be observed in the first years of JWST operations. Our work demonstrates that detailed atmospheric measurements of these planets will also allow us to characterise their internal structures in detail. This new era of high-quality sub-Neptune observations will therefore usher in unprecedented insight into the inner workings of these mysterious worlds.

\acknowledgments

We thank the anonymous referee for their comments, which improved the quality of this manuscript. The NASA/CSA/ESA JWST observations shown in this work are associated with the GO program \#1803 (PI: Kempton). Data were obtained from the Mikulski Archive for Space Telescopes at the Space Telescope Science Institute, which is operated by the Association of Universities for Research in Astronomy, Inc., under NASA contract NAS 5-03127 for JWST. The specific observations analyzed can be accessed via \dataset[DOI: 10.17909/qe3z-qj40]{https://doi.org/10.17909/qe3z-qj40}. Support for this program was provided by NASA through a grant from the Space Telescope Science Institute. This research was also supported by the AEThER program, funded in part by the Alfred P. Sloan Foundation under grant \#G202114194, as well as by NASA ADAP 80NSSC19K1014. M.S. and M.Z. acknowledge support from the 51 Pegasi b Fellowship, funded by the Heising-Simons Foundation. This research has made use of the NASA Astrophysics Data System and the NASA Exoplanet Archive, as well as the Python packages \textsc{NumPy} \citep{Harris2020}, \textsc{SciPy} \citep{Virtanen2020}, and \textsc{Matplotlib} \citep{Hunter2007}.

\vspace{5mm}

\bibliography{references}
\bibliographystyle{aasjournal}

\end{document}